# CONTRAST-ENHANCED X-RAY IMAGING OF ARTICULAR CARTILAGE: RELIABILITY OF A CATIONIC CONTRAST AGENT IN COMBINATION WITH HIGH-RESOLUTION PERIPHERAL QUANTITATIVE COMPUTED TOMOGRAPHY SYSTEM

Simone Fantoni [1], Matteo Berni [1], Roberta Fognani [1], Giulia Fraterrigo [2], Paolo Cardarelli [3], Fabio Baruffaldi [1] and Massimiliano Baleani [1]

1 Laboratorio di Tecnologia Medica, IRCCS Istituto Ortopedico Rizzoli, 40136 Bologna, Italy

2 Laboratorio di BioIngegneria Computazionale, IRCCS Istituto Ortopedico Rizzoli, 40136 Bologna, Italy

3 Istituto Nazionale di Fisica Nucleare (INFN), Division of Ferrara, 44122 Ferrara, Italy

Abstract

Articular cartilage exhibits distinctive mechanical behaviour, attributable to its biphasic composition and hierarchical organization. Proteoglycans, essential constituents of the extracellular matrix, contribute to tissue swelling, stiffness, and viscoelasticity, thanks to their fixed charge density. Degenerative alterations in proteoglycan content and collagen structure compromise the mechanical integrity of articular cartilage. Early detection of these alterations is essential to increase the chance of effectiveness of any innovative treatment. Although magnetic resonance imaging provides valuable compositional information, its limited spatial resolution restricts its effectiveness in capturing subtle alterations in articular cartilage. As an alternative, contrast-enhanced X-ray imaging circumvents such limitation, resorting to the use of radiopaque contrast agents. In this context, cationic contrast agents, like CA4+, enable the quantitative assessment of proteoglycan content via electrostatic attraction. High-resolution peripheral quantitative computed tomography offers an optimal compromise between spatial resolution and radiation exposure, making it a promising tool for clinical use. This study aimed to explore the relationship between proteoglycan content—quantified through contrast-enhanced high-resolution peripheral quantitative computed tomography combined with the use of CA4+—and the viscoelastic properties of healthy bovine articular cartilage, assessed via indentation testing. These preliminary data from a small sample size group support significant correlations between cartilage enhanced radiopacity and parameters representing the tissue's mechanical behaviour. These findings should be confirmed in a larger sample size group.

1. INTRODUCTION

Articular cartilage (AC) is a soft tissue showcasing unique mechanical properties, arising from its composition and hierarchical structure. More precisely, two phases can be distinguished [1]. The solid phase includes collagen network and proteoglycans (PGs), as the main components of the extracellular matrix (ECM). The collagen network is composed by fibres locally collected into bundles, anchored to the calcified cartilage, and expanding towards the articular surface in an arcade-like configuration [2]. PG are macromolecules including glycosaminoglycan chains, featuring a negative charge unbalance, the fixed charge density (FCD) [3]. Intertwined with collagen network, PGs outstretch owing to the mutual repulsion, and distribute along the tissue depth according to a characteristic concentration gradient [2]. The fluid phase, represented by the interstitial fluid, is mainly composed of water and positively-charged ions [4].

Due to the interplay of those hierarchically arranged phases, AC responds to any local stimulus with an instantaneous flow-independent response and a flow-dependent viscoelastic response [5]. The instantaneous response originates from the tensile deformation of collagen bundles in the superficial layer of the tissue, the rearrangement of the collagen network in the deep layer, the mutual repulsion of glycosaminoglycans bound to PGs and fluid pressure. The originated fluid pressure triggers the flow of interstitial fluid across the ECM. Viscous phenomena exhaust when the equilibrium between the mechanical stimulus and the tissue reaction is reached [6]. The FCD of PGs plays a crucial role in pretensioning the collagen network and, moreover, originates the osmotic pressure. The combination of both phenomena induces the swelling of AC [7].

The degradation of AC owed to musculoskeletal diseases, i.e., osteoarthritis, leads to a progressive decline of the tissue mechanical competence [8] Several alterations of AC serve as hallmarks for potential identification of tissue degeneration, among which the loss of PG content implicates a decrease in the tissue stiffness [9]. Moreover, the fibrillation of collagen network induces the augmented permeability of ECM, held accountable for altering both the instantaneous and viscous response of the tissue [10].

The early detection of any alteration of AC could be crucial in the field of regenerative medicine and pharmacological treatments. For instance, the capability to monitor their impact on disease progression could delay the need for invasive solutions such as prosthetic implant [11]. In this frame, diagnostic techniques sensitive to one or more AC components should be considered in the perspective of identifying early degenerative hallmarks. Magnetic resonance imaging (MRI) delivers high-contrast images of hydrated tissues and enables the quantitative assessment of several parameters signalling the tissue condition. For instance, T1ρ and T2 relate with the PG content and state of collagen mesh, respectively [12]. Nevertheless, clinical MRI systems suffer from low spatial resolution, with major implications for the evaluation of thin layers of tissues, as in the case of AC [13].

Conventional X-ray imaging methods have proved their reliability mainly in the evaluation of mineralized tissues [14]. Although such methods can overcome the MRI limitation related to spatial resolution [15], they are not suitable to identify soft tissues' features due to their low electron density.

To enhance the X-ray visibility of soft tissues, the use of radiopaque exogenous substances is required. Contrast agents (CAs) are compounds that include electron-dense elements (e.g., iodine) and are employed in both research and clinical context to increase the radiopacity of a target organ or tissue [16]. Clinical CAs feature chemical properties (e.g., pH and osmolality) meeting strict requirements for patient safety [17]. In the research context, CAs include polyoxometalate-based histological stains, nanoparticles or experimental formulations, adopted in both ex vivo and in vitro studies [18]. Regardless of the CA type, it is the net molecular charge that determines the CA affinity for a specific tissue or component. Focusing on AC, cationic CAs provide superior results in terms of imaging outcome, as demonstrated for CA4+. This formulation is an experimental iodine-based cationic CA that allows the quantitative assessment of PG content in AC thanks to the electrostatic attraction exerted by FCD of glycosaminoglycans [19]. Previous studies have reported a correlation between CA4+ contrast enhancement and tissue properties, namely stiffness, compressive modulus, and torsional coefficient of friction [20,21]. Therefore, contrast enhanced X-ray imaging protocols implemented with CA4+ could predict the mechanical competence of AC, providing essential outcomes in both preclinical and clinical contexts. In deploying such a framework clinical X-ray imaging systems should be privileged, prioritizing the acquisition of patient limbs at acceptable spatial resolutions and low radiation doses. High-resolution peripheral quantitative computed tomography (HR-pQCT) scanners represent an optimal trade-off between relatively high spatial resolution (i.e., from 40 to 60 µm) and minimal radiation dose [22]. Nevertheless, insights about the deep connection between the composition and mechanical competence of the tissue must be provided first, e.g., by focusing on ex vivo experiments. The aim of this study was to investigate the relationship between information related to the PG content of healthy bovine AC – retrieved by a contrast-enhanced HR-pQCT clinical protocol – and the tissue viscoelastic properties, derived from indentation test.

2. MATERIALS AND METHODS

2.1. Contrast Agent
The CA employed by this study is CA4+, i.e., with formulation (5,50-[Malonylbis(azanediyl)]bis[N1,N3-bis(2-aminoethyl)-2,4,6-triiodoisophthalamide] chloride). One molecule of CA4+ includes six iodine atoms, and displays iodine fraction per molecule $f = 0.5084$, net positive charge $q = +4$, molecular weight $mw = 1354$ g/mol. CA4+ salts synthesis was performed according to literature [19]. CA solution was prepared by dissolving CA4+ salts in phosphate-buffered saline (PBS) solution (1X, 7.4 pH, Life Technologies Europe B.V., Bleiswijk, The

Netherlands) until reaching a concentration of 10 mgI/mL, effective to enhance significantly the X-ray attenuation of AC [23]. The pH of the CA solution was balance by adding NaOH 4M, i.e., up to a value of 7.4. The osmolarity of PBS and CA solution was measured with an osmometer (TypM 10/25 µL, accuracy 1 mOsm/kg, Löser Messtechnik, Berlin, Germany), yielding 307 mOsm/kg and 419 mOsm/kg, respectively.

2.2. Osteochondral Tissue Samples

Osteochondral (OC) cuboids (N = 11) were excised from tibio-femoral surfaces of N = 6 fresh bovine stifle joints, obtained from a local slaughterhouse within 24 hours post-slaughter for food production purposes. Cuboid samples were obtained through the use of a diamond-coated blade mounted assembled on a circular saw (MDP200 and TR60, respectively; Remet, Italy). The process was performed under continuous water irrigation, to avoid the dehydration and heating of the OC tissues. OC cuboids were then wrapped in gauzes soaked in PBS 1X and frozen at −20 °C (1st freezing cycle). On a different day, OC cuboids were thawed at 4°C in PBS 1X for one hour, after which cylindrical OC cores (10-mm in diameter and height, including AC, SB and TB) were excised by means of a computer numerically controlled milling machine (ProLight 2000, Light Machines Corporation, Manchester, NH, USA) equipped with a diamond-coated coring tool (10 mm inner diameter). Prior to coring, OC cuboids were spatially tilted to retrieve cores with an AC surface as flat as possible. A total of N = 18 OC cores were excised. OC cores were then stored at -20°C in a PBS 1X-soaked gauze until indentation test (2nd freezing cycle).

2.3. Indentation test

OC cores, in subgroups of N = 6, were thawed in PBS 1X at T = 4 °C overnight. The AC thickness was estimated before indentation by means of X-ray imaging. After placing each OC core in a custom-made polymethyl methacrylate sample holder, four planar images (each after a 90–rotation of the core) were acquired via X-ray microCT system (SkyScan 1072, SkyScan, Aartselaar, Belgium) by applying a previously developed protocol [24], i.e., with a pixel size = 11.5 µm. AC thickness was estimated by using a custom-made MATLAB code (MATLAB version R2022b, MathWorks, Natick, MA, USA) [24].

Afterwards, subgroups of N = 6 OC cores were placed in a custom-made polyacetal sample holder with six cavities [24]. Then, PBS 1X (0.5 mL) was poured on the top surface of the OC cores to keep the AC hydrated.

The sample holder was mounted on the testing machine (Mach-1 V500css, Biomomentum Inc., Laval, QC, Canada), and each OC core underwent the following procedure: (i) a preconditioning of the AC, i.e., preliminary indentation, (ii) three indentations, each performed after a resting period of 40 min, (iii) substitution of the PBS solution with 0.5 mL of CA solution, followed by a 22 h rest at a room temperature (22 ± 2 °C) to reach equilibrium of the diffusion phenomena [23].

The indentation test foresaw the application by a 6 mm spherical indenter of a maximum nominal deformation equal to 15% of AC thickness, at a deformation rate of 15%/s, perpendicular to the articular surface and at the centre of each OC core [25]. The nominal deformation was held for 300 s [26] to assess variation of stress over time. This protocol was employed to investigate the following parameters: maximum reaction load ($S_0$), instantaneous elastic modulus ($E_0$), time constant ($\tau$) and the stretching parameter ($\beta$) of the load–time relaxation curve, and equilibrium modulus ($E_{eq}$). Full details regarding the computation of these parameters have been published elsewhere [25].

2.4. HR-pQCT Imaging

Following the exposure to CA solution, the subgroup of N = 6 OC cores was acquired using a HR-pQCT system (XtremeCT II, SCANCO Medical AG, Brüttisellen, Switzerland). After removing the CA solution from the top surface of each OC core, the sample holder was placed within HR-pQCT, i.e., by aligning the OC cores' axial direction to the scanning axis, and the cores were acquired (X-ray tube voltage = 68 kV, current = 1.47 mA, filtration = 1 mm Al + 0.2 mm Cu, voxel size = 60.7 µm, scan time = 10 min). Afterward, OC cores were removed from the samples holder, wrapped in gauzes soaked in PBS 1X, and stored at -20°C. HR-pQCT images were reconstructed using a multithreaded Feldkamp-Davis-Kress reconstruction program [27]. The reconstructed volumes underwent semi-automatic segmentation of contrast-enhanced AC. The segmented AC volumes were further processed by a custom-made Matlab code to estimate i) the AC thickness, and ii) the depth-dependent distribution of CA4+ across such a tissue, i.e., in term of attenuation – CT number (HU) – profile.

2.5. Statistical Analysis

The distribution of AC thickness, assessed by microCT and HR-pQCT, mechanical parameters, and volumetric distribution of CA4+ across AC samples was assessed via Kolmogorov-Smirnov test.

The replicability of measuring AC thickness by two different imaging approaches, i.e., microCT and HR-pQCT, was investigated by Pearson correlation.

Eventual trend, i.e., upward or downward, of $S_0$, $E_0$, $\tau$, $\beta$, or $E_{eq}$ values over the indentations was assessed through the Friedman test ($p < 0.05$).

In absence of trend in the mechanical parameters' values computed starting by the three indentations, their median values were considered to investigate relationships with the volumetric distribution of CA4+ within AC (Spearman correlation, $p < 0.05$).

3. RESULTS

The AC thickness of OC cores computed by microCT- and HR-pQCT approaches was in the range (0.8 ÷ 3.0) mm and (0.6 ÷ 2.7) mm, respectively. Both the approaches provided non-normal distributed values (Kolmogorov-Smirnov test, $p < 0.001$; **Table 1**).

**Table 1.** Distribution of the AC thickness estimated by microCT and HR-pQCT approaches.

| Percentile | AC thickness (mm) | |
|---|---|---|
| | microCT | HR-pQCT |
| 5th | 0.8 | 0.6 |
| 25th | 1.1 | 1.0 |
| 50th | 1.8 | 1.8 |
| 75th | 2.6 | 2.4 |
| 95th | 3.0 | 2.7 |

The replicability of measuring AC thickness by the two imaging approaches was investigated by Pearson correlation, which highlighted a significant and strong ($\rho > 0$, $\rho^2 = 0.98$; $p < 0.001$) relationship between microCT and HR-pQCT estimates.

Concerning the indentation of OC cores, a total of 54 tests were performed (neglecting the preconditioning indentation). The testing procedure proved to be repeatable (see the percentage coefficient of variation calculated for the mechanical parameters in **Table 2**).

**Table 2.** Distribution of the coefficient of variation (CV%) for the mechanical parameters across the three indentations.

| Percentile | $S_0$ | $E_0$ | $\tau$ | $\beta$ | $E_{eq}$ |
|---|---|---|---|---|---|
| 5th | 0.4 | 0.9 | 0.5 | 0.9 | 0.5 |
| 25th | 1.1 | 1.5 | 0.7 | 1.0 | 1.2 |
| 50th | 2.0 | 2.0 | 1.2 | 1.0 | 1.8 |
| 75th | 3.2 | 3.7 | 3.1 | 1.0 | 3.3 |
| 95th | 5.2 | 5.2 | 4.3 | 1.0 | 14.1 |

The range of variability, i.e., minimum and maximum values, of the computed parameters are reported in the following: $S_0$: (0.8÷4.3) N; $E_0$: (0.6÷9.2) MPa; $\tau$: (1.7÷10.3) s; $\beta$: (0.342÷0.499); $E_{eq}$: (0.1÷0.6) MPa.

No significant trends over the three test repetitions were found regardless the parameter.

Considering the obtained results, the median value of the mechanical parameters across the three indentations was analysed for each OC core, with the aim of investigating its correlation with the volumetric distribution of CA4+ within AC.

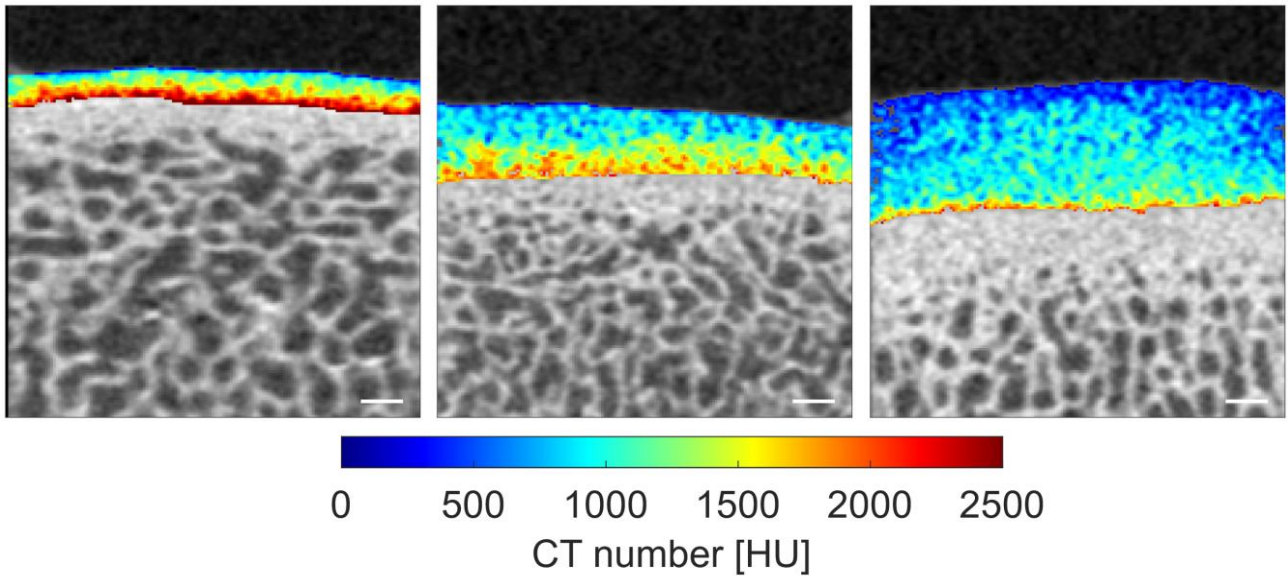

Figure 1. Sections of reconstructed volumes from HR-pQCT acquisitions. Samples representative of thin (left), average (centre) and thick (right) tissue are reported, displaying the radiopacity of contrast-enhanced AC. White line bar = 1 mm.

In the context of HR-pQCT assessment, the range of CA concentration in AC tissue, defined by the minimum and maximum CT numbers, was found to be comparable across OC cores with varying AC thicknesses **(Figure 1)**. Regarding the distribution of CT numbers across the AC thickness, an increasing trend in attenuation values was observed from the articular surface to the deep zone **(Figure 2)**.

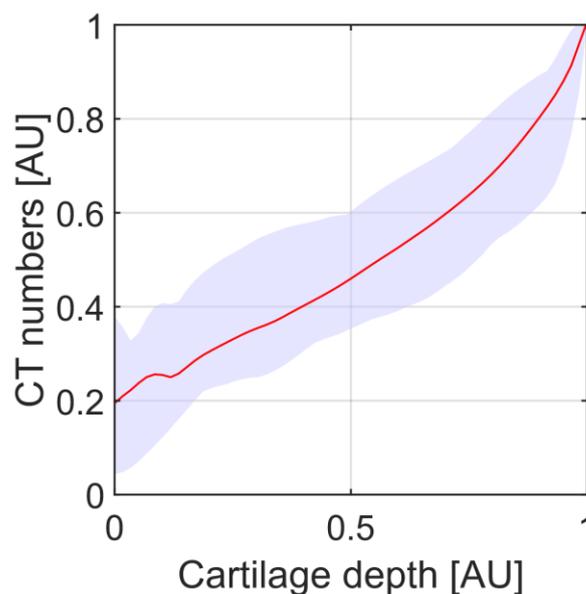

Figure 2. Average profile (red line) of radiopacity, evaluated on N=18 samples previously normalized to their maximum CT number. The average profile points out an increasing radiopacity from the surface (i.e., tissue depth = 0) towards the deep layer (i.e., tissue depth = 1) of AC.

Focusing on the relationship between AC morphology (thickness), composition (PG content), and functionality (viscoelastic properties), several key findings were identified **(Table 3, 4 and 5)**.

Regarding the relationship between AC morphology and composition, a significant negative correlation was observed, specifically between tissue thickness and CT number: $\rho = -0.888$ ($p < 0.001$) **(Table 3)**.

**Table 3.** Spearman's correlation coefficients ($\rho$) between AC morphology – i.e., tissue thickness, computed by HR-pQCT approach – and functionality – i.e., viscoelastic properties. * $p < 0.05$; ** $p < 0.01$; *** $p < 0.001$.

|  | $E_0$ | $S_0$ | $\tau$ | $\beta$ | $E_{eq}$ |
|---|---|---|---|---|---|
| **AC Thickness** | $\rho = -0.926$ *** | $\rho = -0.685$ *** | $\rho = 0.661$ * | $\rho = -0.683$ ** | $\rho = -0.323$ * |

Moving to the relationship between AC composition and functionality, significant, weak to strong, positive correlations were found between CT number and tissue elastic properties (**Table 4**).

**Table 4.** Spearman's correlation coefficients ($\rho$) between AC composition – i.e., PG content, computed by CE HR-pQCT – and functionality – i.e., elastic properties. * $p < 0.05$; ** $p < 0.01$; *** $p < 0.001$.

|  | $E_0$ | $S_0$ | $E_{eq}$ |
|---|---|---|---|
| **CT number** | $\rho = 0.943$ *** | $\rho = 0.752$ *** | $\rho = 0.455$ |

Furthermore, significant, weak to moderate relationships were highlighted between CT number and AC viscous properties (**Table 5**).

**Table 5.** Spearman's correlation coefficients ($\rho$) between AC composition – i.e., PG content, computed by CE HR-pQCT – and functionality – i.e., viscous properties. * $p < 0.05$; ** $p < 0.01$; *** $p < 0.001$.

|  | $\tau$ | $\beta$ |
|---|---|---|
| **CT number** | $\rho = -0.589$ * | $\rho = 0.660$ ** |

## 4. DISCUSSION

The purpose of the present study was to investigate possible correlations between mechanical properties and data achievable from high-resolution X-ray scan of the contrast-enhanced AC.

The findings for AC thickness estimates highlighted consistency between the microCT and HR-pQCT approaches, supporting the reliability of HR-pQCT for evaluating contrast-enhanced AC thickness. Minor discrepancies observed between the two modalities are likely attributable to differences in image acquisition—planar in the case of microCT and volumetric for HR-pQCT—as well as variations in spatial resolution. Specifically, the higher resolution of microCT allows for the detection of finer structural features within the tissue.

The design of the study involved the investigation of the AC mechanical properties prior to performing the protocol required by contrast-enhanced HR-pQCT. This choice was based on the known effect of the contrast agent on the mechanical response of AC [24], which could have introduced bias into the findings of the study.

The findings derived from contrast-enhanced HR-pQCT revealed a depth-dependent distribution of CT numbers within the AC tissue. Owing to the cationic nature of the contrast agent employed in this study, the observed increase in CT number with depth indicates a corresponding progressive increase in PG content across the AC thickness [23]. These results align with previous evidence on PG distribution, as demonstrated by a range of ex vivo reference techniques—including histology [28], histochemistry [29], digital densitometry [30], FT-IR spectroscopy [31], and Raman spectroscopy [32]—as well as imaging modalities such as contrast-enhanced X-ray techniques [19–21,33,33].

The present study highlighted a significant negative correlation between AC thickness and tissue PG content. This suggests that PG play a crucial contribute in supporting compressive loads particularly in healthy, thin AC. This evidence is at odds with the findings of a previous study, which showed an increase in PG as a function of AC thickness [34]. Such a discrepancy might be ascribed to both the nature of the species being investigated and sample size, rather than the specific approach employed to estimate the amount of PG content. The correlations between AC thickness and mechanical properties – i.e., considering the negative correlations with the elastic response support the hypothesis that a high PG content within thin AC provides a stiff response to external, compressive stimuli. Of course, these evidence and hypotheses are limited to healthy AC – as the one herein investigated. They cannot be generalised to pathological tissue, in which a decrease in thickness and PG content results in a progressive deterioration of the tissue mechanical properties as the disease progresses [35–37].

Proteoglycans contribute significantly to the load bearing capacity of AC. Their FCD [38] generates osmotic pressure within the interstitial fluid, which is crucial to withstand external compressive stimuli [1]. By accounting for both electrostatic and non-electrostatic interactions, proteoglycans contribute substantially to the overall modulus of AC. In addition to their role in stiffness, proteoglycans are also involved in the tissue's viscous behaviour, particularly through flow-independent intrinsic mechanical responses of the ECM constituents [39]. AC is composed of biological polymers engaged in heterogeneous and complex interactions. According to such heterogeneity, different polymers and interactions are involved in the diverse phase of viscous phenomenon, i.e., reptation of monodisperse polymers, and transient binding of polydisperse polymers (macromolecules) [40].

In agreement with the available evidence, the results obtained in this study from a small sample support the existence of correlations between CT number (PG content) and AC mechanical properties, considering both elastic and viscous parameters.

Regarding the elastic properties, the positive correlations between CT number and $S_0$, $E_0$, and $E_{eq}$ (see **Table 4**) suggest that the higher the PG content, the stiffer the instantaneous and equilibrium response of the AC. Such evidence are in strong agreements with the literature, considering both the instantaneous [20,41–43] and equilibrium [41,44–46] elastic parameters and regardless the approach employed to estimate the PG content.

Concerning the viscous parameters, correlations were found between CT number and τ or β suggesting that the higher the CT number (PG content), the faster the exhaustion of the viscous phenomena. Due to the complex nature of the phenomena underlying fluid-independent viscosity, the topic deserves further discussion.

The relaxation time τ is linked to the physical features of the polymers composing the system, e.g., concentration, mobility and stiffness. According to the theory of polymer dynamics, faster stress-relaxation (i.e., lower τ values) is associated with an increased molecular mobility and a decreased molecular stiffness [47].This suggests that fluid-independent relaxation processes reach an equilibrium more rapidly in systems composed of more compliant polymers. This hypothesis is supported by the findings of this study, where i) faster relaxation was observed in more compliant AC, as evidence by a significant negative correlation between τ and $E_0$ (ρ = – 0.544, p < 0.05), and ii) thicker tissue exhibited a more compliant mechanical response. In summary, the following trend was observed: increased AC thickness is associated with lower CT numbers (i.e., lower PG content), which in turn suggests higher molecular mobility of polymers, a more compliant tissue response, and consequently, a more rapid dissipation of viscous phenomena.

The stretching parameter β depends on the type of polymers composing AC and on their relative motion [37]. A previous study suggested that a decrease in *β* is associated to an increase in the molecular weight of the entangled polymers [48]. The positive correlation between CT number (PG content) and β herein highlighted is supported by existing literature [45].

This study presents several limitations that should be acknowledged. First, no gold standard techniques for the assessment of PG content were employed to validate the data obtained from contrast-enhanced HR-pQCT. Nevertheless, it is important to note that the electrostatic affinity between CA4+ molecules and the fixed negative charges of PG is a well-established and extensively studied phenomenon, consistently confirmed across numerous studies [33,49–52].

Second, the time interval between the mechanical characterization of AC via indentation testing and the estimation of PG content through HR-pQCT—approximately 22 hours—may have led to progressive tissue degradation, potentially affecting its compositional integrity. However, this time frame was previously identified as the minimum duration required to achieve equilibrium in CA4+ diffusion within AC [23], and thus represents the necessary condition for reliable volumetric PG quantification.

Third, the sample size of the present study is limited. A larger cohort would be necessary to more accurately characterize and statistically validate the correlations observed between morphological, compositional, and mechanical parameters.

Moreover, the correlations reported in this work pertain exclusively to healthy AC. The onset and progression of joint pathologies, such as osteoarthritis, are known to alter the principal features of AC, notably reducing PG content and modifying its mechanical response [53–56]. Nonetheless, it is worth noting that findings from comparable studies investigating both healthy and pathological AC report consistent results, suggesting that degenerative conditions do not corroborate the fundamental relationships among tissue morphology, composition, and mechanics [45,57]. Despite the above limitations, the present study supports the existence of significant correlations among the morphology, composition and the mechanical response of healthy AC. These findings lay the groundwork for a more comprehensive and integrative evaluation of key cartilage features in both research and clinical contexts.

5. CONCLUSIONS

Contrast-enhanced X-ray imaging allows for the identification of a significant correlation between AC thickness, PG content, and its viscoelastic mechanical properties. This approach holds considerable potential for the prediction of critical tissue characteristics in future clinical contexts. The methodologies introduced in this study may contribute to improving the early diagnosis of AC degeneration. Furthermore, they offer a valuable tool for assessing the effectiveness of therapeutic strategies and biomaterials aimed at AC repair and regeneration. Ultimately, these methods could inform and refine tissue engineering strategies, supporting the development of scaffolds that more accurately replicate the mechanical behaviour of native AC.